\documentclass[11pt]{article}
\usepackage[normalem]{ulem}
\usepackage{hyperref}
\usepackage{amsmath}
\usepackage{braket}
\usepackage{amsthm}
\usepackage{amssymb}
\usepackage{url}
\usepackage{algorithm}
\usepackage{algpseudocode}
\usepackage{mathtools}
\usepackage[title]{appendix}
\usepackage[all]{xy}
\usepackage{geometry}
\usepackage{physics}
\usepackage{cancel}
\usepackage{mdframed}
\usepackage[most]{tcolorbox}
\usepackage{enumitem}
\usepackage{xcolor}
\usepackage{comment}
\usepackage{array}
\usepackage{stackengine,scalerel}
\usepackage{cleveref}
\usepackage{csquotes}

\newtheorem{theorem}{Theorem}[section]

\newtheorem{lemma}[theorem]{Lemma}
\newtheorem{proposition}[theorem]{Proposition}

\newtheorem*{conjecture*}{\textbf{Conjecture}}

\newtheorem{claim}{Claim}[theorem]

\theoremstyle{definition}
\newtheorem{definition}{Definition}

\newtheorem{remark}[theorem]{Remark}
\newtheorem*{remark*}{\textbf{Remark}}

\newcommand{\negl}{\mathsf{negl}}
\newcommand{\Gen}{\mathsf{Gen}}
\newcommand{\Commit}{\mathsf{Commit}}
\newcommand{\Open}{\mathsf{Open}}
\newcommand{\Ver}{\mathsf{Ver}}
\newcommand{\Out}{\mathsf{Out}}

\newcommand{\pk}{\mathsf{pk}}
\newcommand{\sk}{\mathsf{sk}}
\newcommand{\eps}{\varepsilon}
\newcommand{\poly}{\text{poly}}

\newcommand{\secp}{\lambda}
\newcommand{\QPT}{\mathsf{QPT}}
\newcommand{\QFHE}{\mathsf{QFHE}}
\newcommand{\brho}{\boldsymbol{\rho}}
\newcommand{\bsigma}{\boldsymbol{\sigma}}
\newcommand{\bphi}{\boldsymbol{\phi}}
\newcommand{\PPT}{\mathsf{PPT}}

\newcommand{\calI}{\mathcal I}

\newcommand{\calL}{\mathcal L}

\newcommand{\sfA}{\mathsf A}
\newcommand{\sfB}{\mathsf B}
\newcommand{\sfC}{\mathsf C}
\newcommand{\sfD}{\mathsf D}

\newcommand{\sfO}{\mathsf O}

\newcommand{\sfQ}{\mathsf Q}
\newcommand{\sfR}{\mathsf R}

\newcommand{\bbN}{\mathbb N}

\newtheorem{protocol}[theorem]{Protocol}

\Crefname{importedtheorem}{Imported Theorem}{Imported Theorems}
\Crefname{theorem}{Theorem}{Theorems}
\Crefname{proposition}{Proposition}{Propositions}
\Crefname{claim}{Claim}{Claims}
\Crefname{lemma}{Lemma}{Lemmas}
\Crefname{conjecture}{Conjecture}{Conjectures}
\Crefname{corollary}{Corollary}{Corollaries}
\Crefname{construction}{Construction}{Constructions}
\Crefname{property}{Property}{Properties}

\theoremstyle{definition}

\Crefname{definition}{Definition}{Definitions}
\Crefname{assumption}{Assumption}{Assumptions}
\Crefname{notation}{Notation}{Notations}

\theoremstyle{remark}

\Crefname{question}{Question}{Questions}
\Crefname{remark}{Remark}{Remarks}
\Crefname{comment}{Comment}{Comments}
\Crefname{fact}{Fact}{Facts}

\newcommand{\aux}{\mathsf{aux}}

\newcommand{\MF}{\mathsf{MF}}

\newcommand{\BQP}{\mathsf{BQP}}

\newcommand{\LWE}{\mathsf{LWE}}

\newcommand{\Ext}{\mathsf{Ext}}

\newcommand{\Real}{\mathsf{Real}}

\newcommand{\Ideal}{\mathsf{Ideal}}

\newcommand{\V}{\mathsf{V}}
\newcommand{\QIP}{\mathsf{QIP}}
\newcommand{\QIA}{\mathsf{QIA}}
\newcommand{\MIP}{\mathsf{MIP}}

\def\cC{{\mathsf{C}}}
\def\cV{{\mathsf{V}}}
\def\cP{{\mathsf{P}}}
\def\D{\mathsf D}
\def\S{\mathsf S}

\def\Measure{\mathsf{Measure}}

\newcommand{\QMA}{\mathsf{QMA}}
\newcommand{\QMATIME}{\mathsf{QMATIME}}
\newcommand{\QPTIME}{\mathsf{QPTIME}}
\newcommand{\QTIME}{\mathsf{QTIME}}

\newcommand{\PRG}{\mathsf{PRG}}

\newcommand{\btau}{\boldsymbol{\tau}}

\newcommand{\bb}{\boldsymbol{b}}
\newcommand{\bm}{\boldsymbol{m}}
\newcommand{\bs}{\boldsymbol{s}}
\newcommand{\bv}{\boldsymbol{v}}
\newcommand{\by}{\boldsymbol{y}}
\newcommand{\bz}{\boldsymbol{z}}
\newcommand{\bzero}{\mathbf{0}}
\newcommand{\bone}{\mathbf{1}}

\DeclareMathOperator*{\E}{\mathbb{E}}

\newlist{caselist}{enumerate}{1}
\setlist[caselist]{label=\arabic*., ref=\arabic*}
\crefname{caselisti}{Case}{Cases}

\title{Compiling Any $\MIP^*$ into a (Succinct) Classical Interactive Argument}
\date{}
\author{Andrew Huang \\ MIT CSAIL \\ \texttt{ahuang@mit.edu} \and Yael Tauman Kalai \\ MIT CSAIL \\ \texttt{tauman@mit.edu}}
\begin{document}
\maketitle
\begin{abstract}
    We present a generic compiler that converts any $\MIP^*$ protocol into a \emph{succinct} interactive argument where the communication and the verifier are \emph{classical}, and where post-quantum soundness relies on the post-quantum sub-exponential hardness of the Learning with Errors ($\LWE$) problem. Prior to this work, such a compiler for $\MIP^*$ was given by Kalai, Lombardi, Vaikuntanathan and Yang (STOC 2022), but the post-quantum soundness of this compiler is still under investigation.
    
    More generally, our compiler can be applied to any $\QIP$ protocol which is sound only against semi-malicious provers that follow the prescribed protocol, but with possibly malicious initial state.  Our compiler consists of two steps. We first show that if a language $\calL$ has a $\QIP$ with semi-malicious soundness, where the prover runs in time $T$, then $\calL \in \QMATIME(T)$. Then we construct a succinct classical argument for any such language, where the communication complexity grows polylogarithmically with $T$, under the post-quantum sub-exponential hardness of $\LWE$.

    \textbf{Note: After this work was submitted, an independent and concurrent work \cite{BKL+25} resolved the question of quantum soundness of the KLVY compiler.}
\end{abstract}
\newpage

\enlargethispage{1cm}
\tableofcontents
\pagenumbering{roman}
\newpage
\pagenumbering{arabic}

\section{Introduction}\label{sec:intro}
Proof systems lie at the foundation of both modern cryptography and complexity theory, and underlie the very definition of the complexity class $\mathsf{NP}$. Since the 1980s, there has been remarkable progress in their study, leading to a sequence of increasingly powerful models: interactive proofs ($\mathsf{IP}$) \cite{GMR85, Bab85}, interactive arguments ($\mathsf{IA}$)~\cite{BCC88} where soundness is guaranteed only against computationally-bounded provers, multi-prover interactive proofs ($\mathsf{MIP}$)~\cite{BOGKW88}, probabilistically checkable proofs ($\mathsf{PCP}$)~\cite{AS98, ALMSS98}, and more recently, interactive oracle proofs ($\mathsf{IOP}$)~\cite{BCS16, RRR16}. These notions have been deeply influential in both theory and practice, yielding breakthroughs such as the PCP Theorem and inapproximability results, and enabling scalability in blockchain and other cryptographic applications.

The advent of quantum computation has motivated the study of various proof models in the quantum setting. Quantum interactive proofs ($\QIP$)~\cite{Wat99} generalize classical interactive proofs by allowing both prover and verifier to be quantum machines exchanging quantum messages. Similarly, quantum multi-prover interactive proofs ($\mathsf{QMIP}$)~\cite{KM03} generalize classical multi-prover interactive proofs by allowing all parties to be quantum machines exchanging quantum messages. One other class of multi-prover protocols which has been studied is $\MIP^{*}$, where the provers are quantum and may share entanglement, but where the verifier (and communication) is restricted to being classical \cite{CHTW04}. The complexity classes $\mathsf{QMIP}$ and $\MIP^{*}$ are known to be equivalent \cite{RUV13}, and in fact the conversion between the two classes of protocols preserves the computational efficiency of honest provers (assuming the use of two additional provers and more rounds).

In both the classical and quantum settings, the overarching goal of a proof system is for a computationally powerful prover (or provers) to convince a computationally bounded verifier of the truth of a statement. Typically, the computational power refers to the time complexity, where often the verifier is assumed to run in polynomial time, while the prover(s)  may run in super-polynomial time. However, time complexity is not the only axis of separation: a prover (or provers) may also have access to quantum capabilities that the verifier lacks. We will refer to protocols with a single {\em quantum} prover but {\em classical} polynomial-time verifier as classical $\mathsf{QIP}$'s (or classical $\mathsf{QIA}$'s in the case where soundness holds only against computationally bounded provers). 

An important question that received significant attention in recent years is the following: 

\begin{quote}
\centering \itshape Which quantum computations can we classically verify (efficiently and succinctly)? 
\end{quote}

\subsection{Prior Work}
A notable milestone in the study of classical $\mathsf{QIA}$'s was due to Mahadev~\cite{Mah18}, who constructed a classical $\mathsf{QIA}$ for $\mathsf{QMA}$ based on the post-quantum hardness of $\LWE$, enabling a purely classical verifier to validate any $\QMA$ statement. In an alternative line of work, Kalai, Lombardi, Vaikuntanathan, and Yang \cite{KLVY23} introduced a compiler to convert any $\MIP^{*}$ protocol into a classical $\QIA$ assuming $\QFHE$. However, they only proved the soundness of the resulting classical $\QIA$  against classical (computationally-bounded) malicious provers. 

In a subsequent work, Natarajan and Zhang \cite{NZ23} analyzed the post-quantum soundness of the KLVY compiler when applied to a specific nonlocal game, thus giving a different construction of classical arguments for $\BQP$. Shortly after, Metger, Natarajan, and Zhang \cite{MNZ24} proved the post-quantum soundness of the KLVY compiler when applied to a different \emph{question-succinct} nonlocal game for $\QMA$, thus constructing the first succinct classical arguments for $\QMA$.\footnote{There are known cryptographic techniques for generically compressing answers in question-succinct protocols \cite{LMS22, BKL+22}.}$^,$\footnote{Succinct classical arguments for $\QMA$ were concurrently constructed in \cite{GKNV24} using different techniques, which are relevant to this work, and which we elaborate on in \Cref{sec:overview}.} 

More recent results \cite{KMP+25, KPR+25, BLJS25, BKL+25} have related the soundness of the KLVY-compiled protocol to the commuting operator/quantum value of the underlying nonlocal game. We note that the KLVY-compiled protocol is not question-succinct if the underlying $\MIP^{*}$ is not question-succinct.

\subsection{Our Results}

In this work, we present an alternative compiler which generically transforms any $\MIP^{*}$ (even in the non-succinct or multi-prover setting) for a language into a fully succinct classical $\QIA$. The soundness of our resulting classical $\QIA$ is related to the quantum value of the underlying $\MIP^{*}$, assuming the post-quantum hardness of the $\LWE$ assumption.

Our transformation is prover-efficient if the prover is given multiple copies of the (honest) auxiliary input state of all provers in the underlying $\MIP^*$ and this state is \emph{real-valued} (i.e., of the form $\sum_x \alpha_x \ket{x}$ where $\alpha_x \in \mathbb{R}$ for every $x$).\footnote{The restriction to real-valued auxiliary states is not necessary for soundness and is used only to ensure efficient prover runtime.}  We note that all known protocols for classically verifying quantum computations (starting with Mahadev~\cite{Mah18}) are only prover-efficient under these restrictions.    

The transformed protocol achieves the standard computational soundness guarantee: no cheating prover running in quantum time $\poly(T)$, where $T$ is a parameter $T$ that influences our computational hardness assumption, can convince the verifier of a false statement. Moreover, the resulting classical $\mathsf{QIA}$ is \emph{succinct}: its communication complexity is polynomial in the security parameter $\secp$, and the verifier runs in time $\poly(\secp) + \widetilde{O}(|x|)$, where $x$ is the statement being verified. The prover’s runtime grows only polynomially relative to that of the original protocol.

\begin{theorem}[Informal]\label{theorem:MIP*-compiler}
Let $(P_1, \ldots, P_k, V)$ be any $k$-prover $\MIP^*$ protocol for a language $\calL$. Denote the honest provers' runtime by $t_P$, and suppose the auxiliary states of the honest provers are real-valued. Let $T = T(\secp)$ be any function such that $t_P \leq T(\secp)\leq 2^\secp$.

Assuming the post-quantum $T$-hardness of the Learning with Errors ($\LWE$) problem, there exists a $T$-secure \emph{classical} $\mathsf{QIA}$ for $\mathcal{L}$, where the number of rounds and the communication complexity are $\poly(\secp$), the verifier runtime is $\poly(\secp) + \widetilde{O}(|x|)$, where $x$ is the statement being verified, and the prover runtime is $\poly(t_P)$ (given $\poly(t_P)$ copies of the provers' initial state).
\end{theorem}

More generally, we extend the result above to any $\QIP$ that is sound only against {\em semi-malicious} provers, which are provers that follow the prescribed protocol honestly but with potentially malicious auxiliary input states.
We note that every $\MIP^*$ protocol (even those with only semi-malicious soundness) can be trivially converted into a $\QIP$ with semi-malicious soundness. Specifically, the $\QIP$ prover's initial state can be thought of as the combined initial states of all players in the $\MIP^{*}$; the $\QIP$ verifier sends all of the $\MIP^{*}$'s referee's queries to the prover (in the clear), and the $\QIP$ prover sends the answers obtained by emulating the players in the $\MIP^*$ protocol. Semi-malicious soundness holds since the malicious prover must act honestly according to the (local) prescribed actions of the $\MIP^*$ provers.\footnote{Our protocol also readily applies to $\mathsf{QMIP}$'s with semi-malicious soundness for the same reason.} Thus, our work extends the classical-verifier paradigm to a broader class of quantum proof systems.

We state this more general result below:
\begin{theorem}[Informal]\label{theorem:compiler}
Let $(P, V)$ be any semi-malicious $\QIP$ for a language $\calL$ with prover runtime $t_P$, and suppose the auxiliary state $\ket{\psi}$ of the honest prover is real-valued. Let $T = T(\secp)$ be any function such that $t_P \leq T(\secp)\leq 2^\secp$.

Assuming the post-quantum $T$-hardness of the Learning with Errors ($\LWE$) problem, there exists a $T$-secure \emph{classical} $\mathsf{QIA}$ for $\mathcal{L}$, where the number of rounds and the communication complexity are $\poly(\secp$), the verifier runtime is $\poly(\secp) + \widetilde{O}(|x|)$, where $x$ is the statement being verified, and the prover runtime is $\poly(t_P)$ (given $\poly(t_P)$ copies of $\ket{\psi}$).
\end{theorem}

As a building block for proving Theorem \ref{theorem:compiler}, we rely on semi-succinct commitment schemes, which were first defined and constructed in~\cite{GKNV24}. We elaborate on how this is done in \Cref{sec:overview}.

\paragraph{Open Problems.}
Our result raises some intriguing questions related to classical verification of quantum computation. First, is the restriction to real-valued witnesses necessary for prover efficiency? We remark that this is an open problem even for $\QMA$, as the protocols of \cite{Mah18, MNZ24, GKNV24} (implicitly) established verification only for languages with this restriction. While the \cite{KLVY23} compiler presents a different potential route to classical verification, current applications to $\QMA$ verification \cite{MNZ24} also employ a reduction to a $\QMA$-complete language which implicitly assumes real-valued witnesses (or copies of the witnesses for the new language).

Secondly, like all other known classical verification procedures, our protocol is privately verifiable. One tantalizing question is whether this property is inherent. Concretely, for example, could one make our protocol (and others) publicly verifiable with the additional assumption of post-quantum indistinguishability obfuscation ($\mathsf{iO}$)?

Lastly, we believe that similar techniques can be used to analyze a similar compilation for $\MIP^{*}$ \emph{games} (which are not for deciding languages) that preserves the quantum value of the underlying game, as well as to the problem of succinctly sampling from $\mathsf{SampBQP}$. We leave further exploration of these questions for future work.

\section{Technical Overview}\label{sec:overview}
In this section, we highlight the main ideas in the proof of our main result, that any semi-maliciously sound $\mathsf{QIP}$ $(P, V)$ for a language $\mathcal{L}$ can be transformed into a \emph{succinct} classical $\mathsf{QIA}$ for $\mathcal{L}$. We prove this theorem via the following two steps:  

\paragraph{Step 1:}  We first argue that if a language $\calL$ has a semi-malicious $\QIP$ $(P, V)$, where the (honest) prover $P$ runs in time $t_p$, then $\calL\in \QMATIME(t_P)$, where $\QMATIME(t_P)$ is the class of all languages for which membership can be verified by a quantum computer running in time~$\poly(t_P)$.

To this end, we consider the large (uniform) quantum circuit $C$, that takes as input an instance $x$ and the prover's (potentially malicious) auxiliary state $\brho$, and implements an honest interaction between the quantum prover and verifier in the underlying $\QIP$ $(P(\brho), V)(x)$. Specifically, $C$ will use separate registers to simulate the prover's internal state and the verifier's internal state.  It will also use separate designated registers to store the messages passed between the prover and verifier. It will apply the honest prover and verifier's unitaries in alternating order (generating random verifier coins by using ancilla qubits) before applying the verifier's verdict on the state. By flattening the interaction of the $\QIP$ so that $C$ consists only of honest unitaries, we now have a circuit (with public description) whose output behavior can be based on the relaxed notion of semi-malicious soundness.

In particular, $C$ can now be thought of as a $\QMATIME(t_P)$-size verification circuit, where for every $x \in \calL$ its corresponding witness is the auxiliary state $\ket{\psi_x}$ with which the prover $P$ convinces the verifier $V$ that $x\in\calL$, thus establishing that indeed $\calL\in\QMATIME(t_P)$. Importantly, the $\QMATIME(t_P)$ witness for $x \in \calL$ is precisely the auxiliary input $\ket{\psi_x}$ that the prover $P$ uses in the underlying $\QIP$.

\paragraph{Step 2:}

Next, we construct a succinct classical $\QIA$ for every $\calL \in \QMATIME(t_P)$ for which the witness state is real-valued.
Recall that Mahadev \cite{Mah18} constructed a succinct classical $\QIA$ for a $\QMA$-complete language (the Local Hamiltonian language), which happens to have the property that every $x \in \calL$ has a real-valued witness state. Mahadev's proof relies on two major components: 
\begin{itemize}
    \item The Morimae-Fitzsimons protocol \cite{MF16} for $\BQP$ verification which shows how to efficiently convert any $\BQP$ computation into one that can be verified by measuring each qubit only in the $X$ or $Z$ basis.
    \item  A ``weak commitment scheme'' (also known as a ``measurement'' protocol), which she defines and constructs.  Such a scheme allows a committer to generate a classical ``weak commitment'' to any quantum state, in a way that the  committer can later send a classical opening to each qubit in the $X$ or $Z$ basis, with the guarantee that the opening is  consistent with a quantum state, and binding is guaranteed to hold in the $Z$ basis but not necessarily in the $X$ basis.
\end{itemize}
We observe that the Morimae–Fitzsimons transformation generalizes to every language $\calL \in \QMA$ for which every $x \in \calL$ has a real-valued witness state,\footnote{This was already observed by Mahadev~\cite{Mah18} for the specific $\QMA$-complete language $2$-Local Real Hamiltonian.} and moreover to every language $\calL \in \QMATIME(t_P)$ with real-valued witness states. Namely, using this transformation, one can convert any witness for $x\in\calL$, for any $\calL\in\QMATIME(t_P)$, into a witness $\ket{\psi_x}$ that can be verified by measuring each qubit in either the $X$ or $Z$ basis.

At this stage, one could apply Mahadev's protocol, resulting in a classical $\QIA$, where the prover first uses a ``weak commitment'' scheme to commit to this $X/Z$ witness state $\ket{\psi_x}$, the verifier then sends the $X/Z$ measurement bits, and the prover finally sends the desired openings.  This protocol has the desired completeness and soundness guarantees; however, the resulting classical $\QIA$ will have communication complexity and verifier runtime $\poly(t_P)$, which is prohibitively large.

To reduce communication complexity, we use ideas from the work of \cite{GKNV24}, which constructs a \emph{succinct} classical $\QIA$ for the same $\QMA$-complete language as Mahadev. Specifically, we show that the work of \cite{GKNV24} can be generalized to construct a succinct classical $\QIA$ for any language in $\QMATIME(t_P)$. 

This is done by using a \emph{semi-succinct} commitment scheme,\footnote{A ``weak commitment'' is sufficient here, but we need succinctness to reduce the communication and verifier runtime.} as was defined and constructed under the $\LWE$ assumption in \cite{GKNV24}. In this commitment scheme the verifier sends a single {\em succinct} commitment key to the prover (unlike in Mahadev's protocol, which requires a separate key for each qubit). Using such a semi-succinct commmitment scheme we construct a semi-succinct classical $\QIA$ for any $\calL \in \QMATIME(t_P)$, where the verifier's messages are succinct but the prover's messages are of length $\poly(t_P)$. We mention that to do this we need to shrink the verifier's second message of Mahadev's protocol, where the verifier requests $X/Z$ openings. This is done, as in \cite{BKL+22}: by relying on \cite{ACGH20}, one can assume that the verifier's second message is truly random (i.e., that the verifier simply measures each qubit of state in a random $X$ or $Z$ basis).  Then, we can use a pseudorandom generator and send only the short seed to the prover. This results in a semi-succinct classical $\QIA$.

Finally, we use the protocol compression technique of~\cite{BKL+22} to transform the above semi-succinct classical $\QIA$ into a \emph{fully succinct} one, thereby achieving the target communication complexity $\poly(\secp)$ and verifier runtime $\poly(\secp) + \widetilde{O}(|x|)$. 

We mention that the final protocol is a bit more complicated. First, when applied to $\QMATIME(t_P)$, the Morimae-Fitzimons compiler results in a completeness-soundness gap of only $\frac{1}{\poly(t_P)}$. Therefore, we need to run our protocol the prover needs to use $\poly(t_P)$ copies of the initial state, convert each copy to a history state and commit to all of them. Second, Mahadev's protocol (as well as the succinct version described above) only has (computational) soundness $1-\frac{1}{\poly(\secp)}$. This is the case since soundness is guaranteed only if the cheating prover succeeds in opening the commitment with high probability, in which case we can extract a witness from him. To obtain negligible soundness, we repeat this protocol $\poly(\secp)$ times. While a parallel repetition theorem was proven for Mahadev's protocol \cite{ACGH20, CCY20}, such a result is not known for the succinct version or for private-coin protocols in general (or for KLVY-compiled protocols); thus, we need to repeat our protocol sequentially.

\section{Preliminaries}\label{sec:preliminaries}
\subsection{Concentration Inequalities}
We use the following Chernoff bounds:
\begin{proposition}[Additive Chernoff bound]\label{proposition:chernoff}
Let $X_1, \ldots, X_n$ be i.i.d.\ Bernoulli random variables with expectation $p$. Then for any $\eps > 0$,
    \[ \Pr[\frac{1}{n} \sum_{i=1}^n X_i \leq p - \eps] \leq e^{-2\eps^2n}, \]
and
    \[ \Pr[\frac{1}{n} \sum_{i=1}^n X_i \geq p + \eps] \leq e^{-2\eps^2n}. \]
\end{proposition}

\subsection{Quantum Computation}\label{sec:prelim:quantum-comp}
We start by defining some notation that is used in this paper:
\begin{itemize}
    \item For any random variables $A$ and $B$ (classical variables or quantum states), we use the notation $A \equiv B$ to denote that $A$ and $B$ are identically distributed, and use $A \stackrel{\eps}\equiv B$ to denote that A and B are $\eps$-close, where closeness is measured with respect to total variation distance for classical variables, trace distance for mixed quantum states, and $\|\cdot\|_2$ distance for pure quantum states.

    \item For every two ensemble of distributions $A = \{A_\secp\}_{\secp \in \bbN}$ and $B = \{B_\secp\}_{\secp \in \bbN}$ and for every $\eta = \eta(\secp) \in [0,1)$ we use the notation $A \stackrel{T, \eta}\approx B$ to denote that for every $\poly(T(\secp))$-size distinguisher $D$ and for every $\secp \in \bbN$,
        \[ |\Pr[D(a) = 1]-\Pr[D(b) = 1]| \leq \eta(\secp) \]
    where the probabilities are over $a \gets A_{\secp}$ and $b \gets B_{\secp}$.

    \item For any $q$-qubit quantum state $\brho$ and string $\bb \in \{0, 1\}^{q}$, we denote by $\bm \gets \Measure(\bb, \brho)$ the result of measuring $\brho$ according to the basis $\bb$, where $\bb_i=0$ corresponds to the $Z$ basis, $\bb_i=1$ corresponds to the $X$ basis.
\end{itemize}

\begin{definition}[Adapted from \cite{Wat12}]
    Let $S \subseteq \Sigma^{*}$ be any set of strings. Then a collection $\{Q_s : s \in S\}$ of quantum circuits is said to be \emph{$T$-time generated} if there exists a $\poly(T)$-time deterministic Turing machine that on every input $s \in S$ outputs a description of $Q_s$.
\end{definition}

\begin{definition}[$\QPTIME$]
     We say that a $T$-time generated family of quantum circuits $\{U_n\}_{n \in \bbN}$, where $U_n$ acts on inputs of length $n$, is in $\QTIME[T]$ if $U_n$ consists of at most $T(n)$ constant-qubit gates. We define $\QPTIME[T] := \cup_{c > 0} \QTIME[T^c]$, and let $\QPT := \QPTIME[n]$.
\end{definition}

\begin{definition}[$\QMATIME$]
    Let $T: \bbN \to \bbN$ be a polynomial-time computable function. We say that a language $\calL$ is in $\QMATIME[T]$ if there exists $V \in \QPT$ and functions $c, s$, referred to as the completeness and soundness parameters, respectively,  such that $c(|x|)-s(|x|) \geq \frac{1}{\poly(T(|x|))}$ with the following properties:
    \begin{enumerate}
        \item \textbf{Completeness.} For every $x \in \calL$, there exists a state $\ket{\psi_x}$ of size $\poly(T(|x|))$ such that 
            \[ \Pr[V(x, \ket{\psi_x}) = 1] \geq c(|x|). \]
        We call $\ket{\psi_x}$ a witness for $x\in\calL$, and refer to any set $\{\ket{\psi_x}\}_{x \in \calL}$, consisting of one witness per instance $x \in \calL$, as a complete set of witness states for $\calL$.
        \item \textbf{Soundness.} For every $x \notin \calL$ and every state $\brho$,
            \[ \Pr[V(x, \brho) = 1] \leq s(|x|). \]
    \end{enumerate}
    We refer to $V$ as a $\QMATIME[T]$ verifier for $\calL$ with completeness $c$ and soundness $s$. Without loss of generality, we assume that any ancilla registers used by $V$ are initialized to $\ket{0}$.
\end{definition}

In what follows, for any $\QMATIME[T]$ verifier $V$ which takes as input $m$-qubit states, any $k \in \mathbb{N}$, and any $p \in [0,1]$, we denote by $V^{p, k}$ the verifier which on input an instance $x$ and a $k \cdot m$-qubit state $\ket{\phi}$, applies the circuit $V(x, \cdot)$ to each $m$-qubit block of $\ket{\phi}$ and outputs $1$ if and only if at least $\lceil{pk\rceil}$ verifier circuits output 1. 

We will use the following lemma, which is a trivial extension of the classic $\QMA$ amplification lemma to $\QMATIME[T]$.
\begin{lemma}[$\QMATIME$ amplification] \label{lemma:qmatime_amp}
    Let $V$ be a $\QMATIME[T]$ verifier for a language $\calL$ with completeness $c$, soundness $s$. Let $\{\ket{\psi_x}\}_{x \in \calL}$ be any complete set of witness states for $\calL$.
    
    Then for every $k\in\mathbb{N}$, the following properties hold: 
    \begin{itemize}
        \item For $x \in \calL$,
            \[ \Pr[V^{(c+s)/2, k}(x, \ket{\psi_x}^{\otimes k}) = 1] \geq 1-e^{-k(c-s)^2/2}. \]
        \item For $x \notin \calL$, for all states $\brho$,
            \[ \Pr[V^{(c+s)/2, k}(x, \brho) = 1] \leq e^{-k(c-s)^2/2}. \]
    \end{itemize}
\end{lemma}

\subsection{Quantum Interactive Protocols}
We consider interactive protocols for a language $\calL$ between two parties. One party, denoted by $V$ (for verifier), outputs a single bit, indicating acceptance or rejection. The other party,  denoted by $P$ (for prover), tries to convince the verifier to accept.

\begin{definition}
    A quantum interactive proof ($\QIP$) for a language $\calL$ with completeness $c$ and soundness $s$ is a quantum interactive protocol $(P,V)$  with the following properties:
    \begin{enumerate}
        \item \textbf{Efficiency.} Both parties are quantum and the communication can be quantum, and both the verifier and prover take as input an instance $x$ while the prover additionally takes as input an auxiliary state $\ket{\aux_x}$.\footnote{In some protocols, the auxiliary input may be empty, i.e. $\ket{\aux_x} = \ket{0}$.} 
        In addition, we assume that the honest prover's unitaries can be computed by a Turing machine that runs in the prover's runtime.
        The verifier's unitaries are required to be in $\QPTIME(|x|)$. We denote by
            \[ \langle{(P(\ket{\aux_x}), V)(x)\rangle} \]
        the output bit of $V$ after interacting with $P$.
        \item \textbf{Completeness.} For every $x \in \calL$,
            \[ \Pr[\langle{(P(\ket{\aux_x}), V)(x)\rangle} = 1] \geq c(|x|). \]
        \item \textbf{Soundness.} For every $x \notin \calL$, cheating prover $P^{*}$, and quantum state $\brho_x$, 
            \[ \Pr[\langle{(P^{*}(\brho_x), V)(x)\rangle}=1] \leq s(|x|). \]
    \end{enumerate}
\end{definition}    
In this work, we also consider $\QIP$s with the following weak notion of soundness, which we refer to as \emph{semi-malicious soundness}:
\begin{definition}
    A $\QIP$ $(P, V)$ has \emph{semi-malicious soundness} $s$ if for every $x \notin \calL$ and state $\brho_x$,
        \[ \Pr[\langle{(P(\brho_x), V)(x)\rangle} = 1] \leq s(|x|). \]
\end{definition}
Note that any $\QIP$ with soundness $s$ also trivially has semi-malicious soundness $s$.

\begin{definition}\label{def:QIA}
    Let $T = T(\secp)$ be a function where $\secp \leq T(\secp) \leq 2^{\secp}$. A $T$-secure quantum interactive argument ($\QIA$) for a language $\calL$ is a quantum interactive protocol $(P, V)$ with the following properties:
    \begin{enumerate}
        \item \textbf{Efficiency.} Both the verifier and prover take as input a security parameter $1^{\secp}$ and an instance $x$, and the prover additionally takes as input an auxiliary state $\ket{\aux_x}$. 
        The verifier $V$'s unitaries are required to be in $\QPTIME(|x|,\secp)$.  We denote by
            \[ \langle{(P(\ket{\aux_x}), V)(1^{\secp}, x)\rangle} \]
        the output bit of $V$ after interacting with $P$.
        \item \textbf{Completeness.} For every $x \in \calL$ and every $\secp\in\mathbb{N}$ such that $|x|\leq 2^\secp$,
            \[ \Pr[\langle{(P(\ket{\aux_x}), V)(1^{\secp}, x)\rangle} = 1] \geq 1-\negl(T(\secp)). \]
           
        \item \textbf{$T$-Soundness.} For every $x \notin \calL$ and every $\secp \in \bbN$ such that $|x| \leq 2^\secp$ and such that the honest prover's unitaries are in $\QPTIME(T(\secp))$, it holds that for every quantum prover $P^{*}$ which runs in time $\poly(T(\secp))$, and every $\poly(T(\secp))$-size state $\brho_x$, 
            \[ \Pr[\langle{(P^{*}(\brho_x), V)(1^{\secp}, x)\rangle}=1] \leq \negl(T(\secp)). \]
    \end{enumerate}
\end{definition}

\begin{remark}
    We note that in the definition of a $\QIP$, the completeness and soundness are parameters $c$ and $s$, while in the definition of a $\QIA$ we set these parameters to be $1-\negl(T)$ and $\negl(T)$, respectively. This is done merely for convenience and without loss of generality, since alternatively, we could have added $1^\secp$ as input to the $\QIP$, indicating that the protocol should be run in parallel $\poly(\log T(\secp), 1/(c-s))$ times and accepted if and only if at least $(c+s)/2$ fraction of executions are accepted. By the (threshold) parallel repetition results for $\QIP$ \cite{KW00}, this pushes the completeness and soundness to $1-\negl(T(\secp))$ and $\negl(T(\secp))$, respectively.
    
    Alternatively, we could have defined a $\QIA$ to have completeness $c$ and $T$-soundness $s$ such that $c-s \geq \frac{1}{\poly(|x|)}$, and derive our current definition by considering a threshold sequential repetition.
    
    With this observation, we note that a $\QIP$ is a strictly stronger notion than a $\QIA$, while a semi-malicious $\QIP$ is strictly weaker.
\end{remark}

\begin{remark}
    Throughout this work, we assume for the sake of simplicity that the honest prover's runtime (in any $\QIP$ or $\QIA$) is at least as large as the verifier's runtime.
\end{remark}

\begin{definition}
    A $\QIA$ $(P,V)$ is said to be a \emph{classical $\QIA$} if the verifier and communication are classical (but the prover is still allowed to be quantum). 
\end{definition}

We next present results that show how to convert any $\QIP$ or $\QIA$ into one where the prover and verifier's circuits consist of only Hadamard and Toffoli gates, assuming the prover's auxiliary state $\ket{\aux_x}$ is real-valued (as defined below):
\begin{definition}
    A pure state $\ket{\psi}$ is \textbf{real-valued} if $\ket{\psi} = \sum_{x} \alpha_x \ket{x}$ for $\alpha_x \in \mathbb{R}$. A mixed state $\brho$ is real-valued if it is a mixture of real-valued pure states.
\end{definition}

Assuming the prover's auxiliary state is real-valued, we can assume that all honest unitaries are implemented by circuits with only Hadamard and Toffoli gates, via the efficient transformation below.
\begin{theorem}[Combining \cite{BV93, Kit97, Shi02}]\label{thm:real-valued}
    Let $C$ be a quantum circuit consisting of $m$ constant-qubit gates. Then for every $\epsilon>0$ there exists a circuit $C'$ consisting of $\poly(m, \log(\frac{1}{\eps}))$ Hadamard and Toffoli gates such that $C'(\brho \otimes \ket{0}) \stackrel{\eps}\equiv C(\brho)$ for all real-valued inputs $\brho$. Moreover, there is a $\QPTIME(m, \log(\frac{1}{\eps}))$ algorithm that takes as input $C$ and $\epsilon$ and outputs $C'$.
\end{theorem}

\subsection{From Quantum Witnesses to \texorpdfstring{$X/Z$}{} Witnesses}
We show how to convert any $\QMATIME[t]$ verifer with real-valued witnesses into a verifier with more restricted quantum capabilities. Namely, the new verifier (for the same language) only measures its input in an $X/Z$ basis and performs classical post-processing to make its decision.
\begin{theorem}[Converting Quantum Witnesses to $X/Z$ Witnesses]\label{thm:XZ_tester}
    Fix any function $t= t(|x|)$. Let $\calL$ be any language with a $\QMATIME[t]$ verifier $V$ that has completeness $c$, soundness $s$, and real-valued witness states $\{\ket{\psi_x}\}_{x \in \calL}$.    
    Then there is a $\poly(t)$-time deterministic classical verification procedure $\V_\MF$ such that the following holds:
    \begin{itemize}
        \item \textbf{Completeness.} For every $x \in \calL$, given $\ket{\psi_x}$, there is a $\poly(t)$-time computable state $\ket{\phi_x}$ such that 
            \[ \Pr_{\bb \gets \{0, 1\}^{\poly(t)}, r \gets \{0, 1\}^{\poly(t)}}[\V_\MF(x, r, \bb, \Measure(\bb, \ket{\phi_x})) = 1] \geq \frac{127}{128}-O\left(\frac{1-c}{\poly(t)}\right), \]
            where $\Measure$ is defined in the beginning of Section~\ref{sec:prelim:quantum-comp}.
        \item \textbf{Soundness.} For every $x \notin \calL$ and any state $\brho_x$,
            \[ \Pr_{\bb \gets \{0, 1\}^{\poly(t)}, r \gets \{0, 1\}^{\poly(t)}}[\V_\MF(x, r, \bb, \Measure(\bb, \brho_x)) = 1] \leq \frac{127}{128}-\Omega\left(\frac{1-\sqrt{s}}{\poly(t)}\right). \]
    \end{itemize}
\end{theorem}
We note that the proof of Theorem \ref{thm:XZ_tester} was shown for the $\QMA$-complete language of $2$-Local Real Hamiltonian with real-valued witnesses \cite{KSV02, MF16, ACGH20}, but we observe that one can  extend the proof to \emph{any} $\QMATIME[t]$ language where the verifier has real-valued witnesses. We defer the proof to Appendix \ref{appendix:XZ_tester}. The proof relies on the following theorem.
\begin{theorem}[Adapted from \cite{MF16}]\label{thm:MF16}
    There is a $\poly(q, n)$-time classical sampling procedure $\S_\MF$ and a deterministic (classical) $\poly(q, n)$-time verification procedure $\D_\MF$ such that for any Hamiltonian $H = \sum d_S S$ acting on~$q$ qubits with $n$ terms, where $d_S$ are real numbers and $S$ is a tensor of the Pauli operators $X$, $Z$ and $I$, and any $q$-qubit quantum state $\ket{\psi}$:
        \[ \Pr[\D_\MF(H, b_1, \ldots, b_q, m_1, \ldots, m_q) = 1] = \frac{1}{2}-\frac{\bra{\psi} H \ket{\psi}}{\sum_S 2|d_S|}, \]
    where the probability is over $(b_1, \ldots, b_q) \gets \S_\MF(H)$ and where $(m_1, \ldots, m_q)$ is distributed by measuring $\ket{\psi}$ according to the the $(b_1, \ldots, b_q)$ basis, where $b_i=0$ corresponds to the $Z$ basis, $b_i = 1$ corresponds to the $X$ basis, and $b_i = \bot$ corresponds to not measuring the $i$'th bit and setting $m_i = \bot$.
\end{theorem}

\subsection{Classical Commitments to Quantum States}\label{sec:com}
We use the primitive of classical commitments to quantum states, as defined and constructed in \cite{GKNV24}. This primitive allows a committer to generate a classical commitment to a quantum state and later open each qubit of the state in either the $X$ or the $Z$ basis, where the opening is also classical. More specifically, the committer is given a public key $\pk$ used to generate the commitment, and the corresponding secret key $\sk$ is used to (classically) decode the opening. For our application, we need $\pk$ to be of size $\poly(\secp)$ and not grow with the size of the quantum state being committed. Such a commitment scheme was referred to as a \emph{semi-succinct commitment scheme} in \cite{GKNV24} (since all the algorithms except the key generation will remain non-succinct). 

\subsubsection{Syntax}\label{sec:syntax:semi-succinct}
\begin{definition}[\cite{GKNV24}]
    A semi-succinct classical commitment scheme for quantum states is associated with algorithms $(\Gen, \Commit, \Open, \Ver, \Out)$ which have the following syntax:
    \begin{enumerate}
        \item $\Gen$ is a $\PPT$ algorithm that takes as input a security parameter $\secp$, and outputs a pair $(\pk, \sk) \gets \Gen(1^{\secp})$, where $\pk$ is referred to as the public key and $\sk$ is referred to as the secret key.
        \item $\Commit$ is a $\QPT$ algorithm that takes as input a public key $\pk$ and a $\ell$-qubit quantum state $\bsigma$ and outputs a pair $(\by, \brho) \gets \Commit(\pk, \bsigma)$, where $\by$ is a classical string referred to as the commitment string and $\brho$ is a quantum state.
        \item $\Open$ is a $\QPT$ algorithm that takes as input a quantum state $\brho$, an index $j \in [\ell]$, and a basis $\bb_j \in \{0,1\}$ (where $\bb_j = 0$ corresponds to measuring the $j$'th qubit in the standard basis and $\bb_j = 1$ corresponds to measuring it in the Hadamard basis). It outputs a pair $(\bz_j, \brho') \gets \Open(\brho, (j, \bb_j))$, where $\bz_j$ is a classical string of length $\poly(\secp)$, referred to as the opening string, and $\brho'$ is the residual state (which is sometimes omitted).
        \item $\Ver$ is a $\PPT$ algorithm that takes a tuple $(\sk, \by, (j, \bb_j), \bz)$, where $\sk$ is a secret key, $\by$ is a commitment string to an $\ell$-qubit quantum state, $j \in [\ell]$, $\bb_j \in \{0,1\}$ is a bit specifying the opening basis, and $\bz$ is an opening string. It outputs $0$ (if $\bz$ is not a valid opening) and outputs~$1$ otherwise. 
        \item $\Out$ is a $\PPT$ algorithm that takes a tuple $(\sk, \by, (j, \bb_j), \bz)$ (as above), and outputs a bit $m \gets \Out(\sk, \by, (j, \bb_j), \bz)$.
    \end{enumerate}
\end{definition}
\begin{remark}
    We extend $\Open$, $\Ver$, and $\Out$ to take as input $(\sk, \by, (J, \bb_J), \bz)$ instead of $(\sk, \by, (j, \bb_j), \bz)$, where $J \subseteq [\ell]$ and $\bb_J \in \{0,1\}^{|J|}$, in which case the algorithms run with input $(\sk, \by, (j, \bb_j), \bz)$ for every $j \in J$. In the case of $\Open$ and $\Out$, the new output is simply the concatenation of the outputs for all $j \in J$, while $\Ver$ outputs 1 if and only if the original algorithm accepted on all inputs $j \in J$.
\end{remark}

\subsubsection{Properties}

\begin{definition}
    A classical (semi-succinct) commitment scheme to a quantum state is said to be $T$-secure if it satisfies Definitions \ref{def:SCQ-semi-succinct-correctness} and \ref{def:binding:semi-succ}, listed below.
\end{definition}    

\begin{definition}[Correctness, \cite{GKNV24}]\label{def:SCQ-semi-succinct-correctness}
    A (semi-succinct) classical commitment scheme is correct if for any $\ell$-qubit quantum state $\bsigma$, and any basis $\bb = (b_1, \ldots, b_{\ell}) \in \{0, 1\}^{\ell}$,
    \begin{equation}\label{semi-succinct-correctness-equivalence}
        \Real(1^{\secp}, \bsigma, \bb) \equiv \bsigma(\bb)
    \end{equation}
    where $\bsigma(\bb)$ is the distribution obtained by measuring each qubit $j$ of $\bsigma$ in the basis specified by $b_j$ (standard if $b_j = 0$, Hadamard if $b_j = 1$), and $\Real(1^{\secp}, \bsigma, \bb)$ is the distribution resulting from the following experiment:
    \begin{enumerate}
        \item Generate $(\pk, \sk) \gets \Gen(1^{\secp})$.
        \item Generate $(\by, \brho) \gets \Commit(\pk, \bsigma)$.
        \item Compute $(\bz, \brho') \gets \Open(\brho, ([\ell], \bb))$.
        \item If $\Ver(\sk, \by, ([\ell], \bb), \bz) = 0$ then output $\bot$.
        \item Otherwise, output $\Out(\sk, \by, ([\ell], \bb), \bz)$.
    \end{enumerate}
\end{definition}

\begin{definition}[$T$-Binding, adapted from \cite{GKNV24}]\label{def:binding:semi-succ}
    A classical (semi-succinct) commitment scheme to a multi-qubit quantum state is $T$-binding if there exists a $\QPT$ oracle machine $\Ext$ such that for any $\poly(T(\secp))$-size quantum circuits $\cC^{*}.\Commit$ and $\cC^{*}.\Open$, any function $\ell = \ell(\secp) \leq \poly(T(\secp))$, any $\ell$-qubit state $\bsigma$, and any basis $\bb = (b_1, \ldots, b_{\ell})$,
    \begin{equation}\label{eqn:binding1:semi-succinct} 
        \Real^{\cC^{*}.\Commit, \cC^{*}.\Open}(\secp, \bb, \bsigma) \stackrel{T, \eta}\approx \Ideal^{\Ext, \cC^{*}.\Commit, \cC^{*}.\Open}(\secp, \bb, \bsigma)
    \end{equation}
    where $\eta = O(\sqrt{\delta})$ and
    \begin{equation}\label{eqn:delta:semi-succinct}
        \delta = \E_{\substack{(\pk, \sk) \gets \Gen(1^{\secp}) \\ (\by, \brho) \gets \cC^{*}.\Commit(\pk, \bsigma)}} \max_{\bb' \in \{\bb, \bzero, \bone\}} \Pr[\Ver(\sk, \by, ([\ell], \bb'), \cC^{*}.\Open(\brho, ([\ell], \bb')))=0].
    \end{equation}
    and where $\Real^{\cC^{*}.\Commit, \cC^{*}.\Open}(\secp, \bb, \bsigma)$ is defined as follows: 
    \begin{enumerate}
        \item $(\pk, \sk) \gets \Gen(1^{\secp})$.
        \item $(\by, \brho) \gets \cC^{*}.\Commit(\pk, \bsigma)$.
        \item Compute $(\bz, \brho') \gets \cC^{*}.\Open(\brho, ([\ell], \bb))$.
        \item If $\Ver(\sk, \by, ([\ell], \bb), \bz) = 0$ then output $\bot$. 
        \item Otherwise, let $\bm = \Out(\sk, \by, ([\ell], \bb), \bz)$.
        \item Output $(\pk, \by, \bb, \bm)$.
    \end{enumerate}
    and $\Ideal^{\Ext, \cC^{*}.\Commit, \cC^{*}.\Open}(\secp, \bb, \bsigma)$ is defined as follows:
    \begin{enumerate}
        \item $(\pk, \sk) \gets \Gen(1^{\secp})$.
        \item $(\by, \brho) \gets \cC^{*}.\Commit(\pk, \bsigma)$.
        \item Let $\btau = \Ext^{\cC^{*}.\Open}(\sk, \by, \brho)$.
        \item Measure $\btau$ in the basis $\bb = (b_1, \ldots, b_{\ell})$ to obtain $\bm \in \{0,1\}^{\ell}$.
        \item Output $(\pk, \by, \bb, \bm)$.
    \end{enumerate}
\end{definition}

\begin{theorem}[\cite{GKNV24}]\label{thm:semi_succinct_commitments}
    Under the post-quantum $T$-hardness of Learning with Errors (LWE), there exists a $T$-secure semi-succinct classical commitment to quantum states.
\end{theorem}

\subsection{Verifier-Succinct Protocol Compilation}
Finally, we use a result from \cite{LMS22, BKL+22} which generically converts any classical $\QIA$ with verifier succinctness into a fully succinct argument:
\begin{theorem}[Adapted from \cite{LMS22, BKL+22}]\label{thm:BKL+22}
    Fix any function $\secp \leq T(\secp) \leq 2^{\secp}$. Suppose there is a $2r$-message $T$-secure classical $\QIA$ $(P, V)$ where the verifier's messages that can be computed in $\poly(\secp)$ time and obliviously to the prover's messages. Let $t_P$ be the prover runtime of this argument system.

    Then, assuming the existence of a $T$-collapsing hash function (which is implied by the post-quantum $T$-hardness of LWE), there is a $4(r+1)$-message $T$-secure classical $\QIA$ $(P', V')$ with prover runtime $\poly(t_P)$, verifier runtime $\poly(\secp)+\widetilde{O}(|x|)$, and communication complexity $\poly(\secp)$.
\end{theorem}
\begin{remark}
    Technically, \cite{BKL+22} proves such a statement for $\secp$-security, but the proof extends easily to the setting of $T$-security and arbitrary verifier runtime (assuming the verifier runtime is at most the prover runtime) by using $T$-hardness of LWE instead.
\end{remark}

\section{Our Compiler}\label{sec:succinct-compiler}
We present our compiler which converts any semi-malicious $\QIP$ into a succinct classical $\QIA$. Notably, the communication complexity of our resulting classical $\QIA$ scales only polynomially with the security parameter; the verifier runtime is polynomial in the security parameter with an additive overhead which scales quasi-linearly with the instance length. The prover's runtime is polynomial in its original runtime.

\begin{theorem}
   \label{thm:qip_compiler}
    Let $(P, V)$ be a semi-malicious $\QIP$ for a language $\calL$ with real-valued auxiliary states, completeness $c$, and semi-malicious soundness $s$. Denote by $t_P$ the number of gates in the honest prover's unitary\footnote{We assume that $t_P\geq |x|$.} and suppose $c(|x|)-s(|x|) \geq \frac{1}{\poly(|x|)}$.
    
    Let $T = T(\secp)$ be any function such that $\secp\leq t_P \leq T(\secp) \leq 2^{\secp}$. Then, assuming the post-quantum $T$-hardness of learning with errors, there is a $T$-secure classical $\QIA$ for $\calL$ with $O(\secp^4)$ rounds, prover runtime $\poly(t_P)$ (when given $\poly(t_P)$ copies of the auxiliary state), verifier runtime $\poly(\secp)+\widetilde{O}(|x|)$, and communication complexity $\poly(\secp)$.
\end{theorem}

We prove Theorem~\ref{thm:qip_compiler} via the following two steps. First, in Section \ref{sec:QIP-to-QMATIME} (Theorem \ref{thm:qip_to_qmatime}), we prove that if a language $\calL$ has a  semi-malicious $\QIP$ $(P,V)$ with honest prover runtime in $\QTIME[t_P]$ then $\calL\in \QMATIME[t_P]$, and the witness for $x\in \calL$ consists of $\poly(|x|)$ many copies of the auxiliary state of the honest prover $P$ in the semi-malicious $\QIP$ for $\calL$.
Then, in Section \ref{sec:QMATIME-to-classical} (Theorem \ref{thm:classical_qia_for_qmatime}), we show how to construct a succinct classical $\QIA$ for any language in $\QMATIME[t_P]$ which has real-valued witness states, extending the result of \cite{GKNV24} beyond $\QMA$. 

\subsection{From Semi-Malicious \texorpdfstring{$\QIP$}{} to \texorpdfstring{$\QMATIME[t_P]$}{}}\label{sec:QIP-to-QMATIME}
We prove that if a language $\calL$ has a semi-malicious $\QIP$ with honest prover in $\QTIME[t_P]$ then $\calL\in \QMATIME[t_P]$.
The idea is simple: A witness for $x\in \calL$ is the auxiliary input of the $\QIP$ prover on input~$x$, denoted by $\ket{\aux_x}$. The $\QMATIME[t_P]$ verifier, on input $\ket{\aux_x}$, appends to it ancilla qubits, and simply applies all the prover and verifier unitaries herself to decide if $x \in \calL$.\footnote{Of course, this will create efficiency issues, but we will address this later.}

\begin{theorem}[Converting a $\QIP$ to $\QMATIME$]\label{thm:qip_to_qmatime}
    Let $(P, V)$ be a semi-malicious $\QIP$ for a language $\calL$ with auxiliary states $\{\ket{\aux_x}\}_{x \in \calL}$, completeness $c$, and semi-malicious soundness $s$. Denote by $t_P$ the number of gates in the honest prover's unitary and assume that $|x| \leq t_P \leq 2^{|x|}$. Suppose $c(|x|)-s(|x|) \geq \frac{1}{\poly(|x|)}$.

    Then there is a $\QMATIME[t_P]$ verifier for $\calL$ with completeness $1-\negl(t_P)$ and soundness $\negl(t_P)$. Moreover, the witness state corresponding to an instance $x$ is $\ket{\aux_x}^{\otimes \poly(|x|)}$.
\end{theorem}
\begin{proof}[Proof of Theorem \ref{thm:qip_to_qmatime}]
We first apply a sequential threshold repetition\footnote{We could have applied parallel threshold repetition and relied on \cite{KW00}, but this is inconsequential as this step only serves as a stepping stone.} to $(P, V)$ with $\poly(\log t_P, |x|)$ repetitions and threshold $(c+s)/2$. We denote the resulting semi-malicious $\QIP$ by $(P_1, V_1)$. By an additive Chernoff bound (see Proposition \ref{proposition:chernoff}), $(P_1, V_1)$ is a semi-malcious $\QIP$ with completeness $1-\negl(t_P)$, semi-malicious soundness $\negl(t_P)$, and a $\poly(\log t_P, |x|) \leq \poly(|x|)$ overhead in round complexity, prover runtime, verifier runtime, and communication complexity. The witness state consists of $\poly(|x|)$ copies of the original auxiliary state.

We next use $(P_1, V_1)$ to prove that $\calL \in \QMATIME[t_P]$ with a verifier with the same completeness and soundness. To this end, for notational convenience, we first convert $(P_1,V_1)$ into a three-message public-coin protocol\footnote{This means the verifier sends only classical uniformly random bits to the prover.}, denoted by $(P_2,V_2)$. This can be done using the works of \cite{KW00, MW05, KKM+07}, which show how to compress any $\QIP$ into a three-message and public-coin one with the same completeness and semi-malicious soundness and only $\poly(|x|)$ overhead in all other parameters (including the number of copies of the auxiliary state).

We assume without loss of generality, that the verdict of $V_2$ is computed by applying an efficient unitary to the registers it received from the prover, denoted by $\sfA$ and $\sf{B}$, and to its ancilla registers, denoted by $\sfD$, and measuring the verdict bit which is stored in a designated ancilla output register, denoted by~$\sfO$. 
        
In other words, we assume that $(P_2,V_2)$ has the following syntax, where we denote the initial registers of $P_2$ by $\sfA\sfB\sfC$, the verifier's randomness register by $\sfR$, and its ancilla registers by $\sfD$ and $\sfO$ (where $\sfO$ is the designated output register).
\begin{protocol} \label{protocol:original}
    On input $x$:
    \begin{itemize}
        \item[$\cP_2 \to \cV_2$:] Apply $U_1(x, \cdot)$ to $\ket{\aux_x}_{\sfA\sfB\sfC}$, resulting in state $\bsigma_{\sfA\sfB\sfC}$. Send $\bsigma_{\sfA}$.
        \item[$\cV_2 \to \cP_2$:] Sample $r \gets \{0, 1\}^{\poly(|x|)}$ and send $\ket{r}_{\sfR}$.
        \item[$\cP_2 \to \cV_2$:] Apply $U_2(x, r)$ to $\bsigma_{\sfB\sfC}$, resulting in state $\brho_{\sfB\sfC}$. Send $\brho_{\sfB}$.
        \item[$\cV_2$:] Initialized the ancilla registers $\sfD$ and the output register $\sfO$ to $\ket{0}$. Apply $V_2(x, r, \cdot)$ to registers $\sfA$, $\sfB$, $\sfD$, and $\sfO$. Measure $\sfO$ in the standard basis to receive a bit $b \in \{0, 1\}$. Accept if $b = 1$ and reject otherwise. 
    \end{itemize}
    \end{protocol}
     \begin{remark}
    By linearity, $U_2$ implicitly allows for a ``controlled'' version which can be applied to a superposition of $r$'s:
    \[ U_2(x, \sum_{r \in S} \ket{r} \otimes \ket{\psi}) = \sum_{r \in S} \ket{r} \otimes U_2(x, r, \ket{\psi}). \]
    The same is true for the verifier's verdict unitary $V_2$.
\end{remark}

Now, consider the following verifier $V'$ which takes as input an instance $x$ as well as a state $\ket{\psi}$ on registers $\sfA$, $\sfB$, and $\sfC$, and does the following: 
\begin{enumerate}
    \item Append the ancilla qubits $\ket{0}_{\sfR\sfD\sfO}$ in registers $\sfR$, $\sfD$, and $\sfO$. 
    \item Apply $\calI_{\sfA\sfB\sfC} \otimes H^{|\sfR|} \otimes \calI_{\sfD\sfO}$.
    \item Apply $U_1(x)_{\sfA\sfB\sfC} \otimes \calI_{\sfR\sfD\sfO}$.
    \item Controlled on $r$ in register $\sfR$, apply $\calI_{\sfA} \otimes U_2(x, r)_{\sfB\sfC} \otimes \calI_{\sfD\sfO}$.
    \item Controlled on $r$ in register $\sfR$, apply $V_2(x, r)$ to registers $\sfA$, $\sfB$, $\sfD$, and $\sfO$.\footnote{If $U_1$, $U_2$, and $V_2$ consist only of Hadamard and Toffoli gates, then one can also implement this circuit with only Hadamard and Toffoli gates.}
\end{enumerate}
By observation, $V'$ acts on $\poly(t_P)$ input qubits and uses $\poly(t_P)$ gates and $\poly(t_P)$ ancilla qubits. By the uniformity conditions placed on the prover and verifier's unitaries, we have that $V' \in \QPT$. On input $(x, \brho_x)$ with zeroes in the $\sfR$, $\sfD$, and $\sfO$ registers, the output of $V'$ is simply the output of a random execution of a prover with auxiliary state $\brho_x$ who applies the honest prover's unitary. The completeness and soundness conditions are thus immediately implied by the completeness and semi-malicious soundness of the underlying $\QIP$, and thus $V'$ is a $\QMATIME[t_P]$ verifier with the desired properties.
\end{proof}

\subsection{From \texorpdfstring{$\QMATIME[t_P]$}{} to Succinct Classical \texorpdfstring{$\QIA$}{}} \label{sec:QMATIME-to-classical}
Next, we show how to construct a succinct classical $\QIA$ for any language in $\QMATIME[t_P]$ which has real-valued witness states, extending the result of \cite{GKNV24} beyond $\QMA$. 

\begin{theorem}[A Succinct Classical $\QIA$ for $\QMATIME$]\label{thm:classical_qia_for_qmatime}
    Fix any function $t_P = t_P(|x|)\geq |x|$ and any language $\calL$ with a $\QMATIME[t_P]$ verifier $V$ with completeness $1-\negl(t_P)$, soundness $\negl(t_P)$, and real-valued witness states $\{\ket{\psi_x}\}_{x \in \calL}$, and which uses $\poly(t_P)$ ancilla qubits set to $\ket{0}$.

    Let $T = T(\secp)$ be any function such that $\secp\leq t_P \leq T(\secp) \leq 2^{\secp}$. Then, assuming the post-quantum $T$-hardness of learning with errors, there is a $T$-secure classical $\QIA$ for $\calL$ with $O(\secp^4)$ rounds, verifier runtime $\poly(\secp)+\widetilde{O}(|x|)$, and communication complexity $\poly(\secp)$. Moreover, given access to $\ket{\psi_x}^{\otimes \poly(t_P)}$, the honest prover runs in $\poly(t_P)$ time.
\end{theorem}

\begin{proof}[Proof of Theorem \ref{thm:classical_qia_for_qmatime}]
Starting with a $\QMATIME[t_P]$ verifier $V$ for a language $\calL$, we construct a classical $\QIA$ $(P',V')$ for $\calL$.
\begin{enumerate}
    \item We first construct a non-interactive $\QIP$, denoted by $(P_1, V_1)$, where the verifier makes only $X$ and $Z$ measurements, and where the communication complexity and verifier runtime in $(P_1,V_1)$ scale with $t_P$.  This is done using Theorem \ref{thm:XZ_tester}.

    \begin{protocol}[$\QIP$ $(P_1, V_1)$]\label{protocol:history} On input $x$, where the prover additionally receives a witness state $\ket{\psi_x}$:
        \begin{itemize}
            \item[$\cP_1 \to \cV_1$:] 
            Prepare and send the state $\ket{\phi_x}$ as defined by Theorem \ref{thm:XZ_tester}. Denote by $N$ the number of qubits in the state $\ket{\phi_x}$, and note that $N = \poly(t_P)$.
            \item[$\cV_1$:] When receiving a state $\brho$ from the prover, do the following:
            \begin{enumerate}
                \item Sample $\bb \gets \{0, 1\}^{N}$.
                \item Compute $\bm \gets \Measure(\bb, \brho)$.
                \item Sample $r \gets \{0, 1\}^{\poly(N)}$
                \item Output $\V_{\MF}(1^{\secp}, x, r, \bb, \bm)$.
            \end{enumerate}
        \end{itemize}
    \end{protocol}
    By Theorem \ref{thm:XZ_tester}, $(P_1, V_1)$ is a non-interactive $\QIP$, where the prover runtime has only polynomial overhead given a witness, with verifier runtime $\poly(t_P)$, and communication complexity $\poly(t_P)$, completeness $c = \frac{127}{128}-\negl(t_P)$, and soundness $s = \frac{127}{128}-\Omega(\frac{1-\negl(t_P)}{\poly(t_P)})$. Thus,
        \[ c-s \geq \Omega\left(\frac{1-\negl(t_P)}{\poly(t_P)}\right) - \negl(t_P) \geq \Omega\left(\frac{1}{\poly(t_P)}\right). \]
    Note that $V_1$ is a $\QMATIME[t_P]$ verifier with a very small completeness-soundness gap.
    
    \item We next consider a threshold parallel repetition of $(P_1, V_1)$, which we denote by $(P_2, V_2)$. Let $p =\poly(t_P)$ denote any polynomial such that $c-s \geq p$.

    \begin{protocol}[$\QIP$ $(P_2, V_2)$] 
        \item[$\cP_2 \leftrightarrow \cV_2$:] For every $i \in [\secp^2 \cdot p^2]$, execute $(P_1, V_1)$ in parallel and let $b_i$ be the verdict of $V_1$ in the $i$th execution.
        \item[$\cV_2$:] Output 1 if and only if $\sum_i b_i \geq (c+s)/2 \cdot \secp^2 \cdot p^2$.
    \end{protocol}

    By Lemma \ref{lemma:qmatime_amp}, $(P_2, V_2)$ is a non-interactive $\QIP$ with completeness $c = 1-\negl(T(\secp))$, soundness $s = \negl(T(\secp))$, prover runtime $\poly(t_P)$, verifier runtime $\poly(t_P)$, and communication complexity $\poly(t_P)$. The prover only needs $\poly(t_P, \secp) = \poly(t_P)$ copies of $\ket{\psi_x}$ to compute $\ket{\phi_x}^{\otimes \secp^2 \cdot p^2}$ in $\poly(t_P)$ time. Since $V_2$ is the threshold parallel repetition of $V_1$, this means that $V_2$ behaves by measuring all qubits of its received state in uniformly random basis $\bb$ to get measurement outcomes $\bm$, sampling randomness $r$, and computing some $\poly(t_P)$-time computable classical verdict on $(1^{\secp}, x, r, \bb, \bm)$. We will refer to this classical verdict function by $\V'_\MF$ in reference to the fact that it is doing a modified version of the $\V_\MF$ verifier.
    
    \item We will now ask the prover not to send copies of the witness state but rather a commitment, which we will then ask to open. To this end, we use the following primitives:
    \begin{itemize}
        \item A $T(\secp)$-secure semi-succinct commitment scheme 
        \[
        (\Gen,\Commit,\Open,\Ver,\Out)
        \]
        as defined in Section \ref{sec:com}; the existence of such a commitment scheme is guaranteed under the $T(\secp)$-hardness of $\LWE$ by Theorem \ref{thm:semi_succinct_commitments}.\footnote{We note that as in \cite{Mah18}, we do not need our commitment scheme to be binding on the Hadamard basis; however, we do need the semi-succinctness property.  }
        \item A $T(\secp)$-secure pseudorandom generator 
        \[\PRG: \{0, 1\}^{\secp} \to \{0, 1\}^{\poly(t_P)};
        \] 
        the existence of such a $\PRG$ is implied by the $T(\secp)$-hardness of $\LWE$ (since we rely on our assumption that $t_P \leq T(\secp)$).
    \end{itemize}
    \begin{protocol}[Classical $\QIA$ $(P_3, V_3)$]\label{protocol:classical_semi_succinct} On input $(1^{\secp}, x)$, where the prover additionally receives copies of $\ket{\psi_x}$:
        \begin{itemize}
            \item[$\cV_3 \to \cP_3$:] Generate commitment key $(\pk, \sk) \gets \Gen(1^{\secp})$. Send $\pk$.
            \item[$\cP_3 \to \cV_3$:] With each copy of the witness state, compute the state $\ket{\phi_x}$ as defined by Theorem \ref{thm:XZ_tester}, and define the state 
                \[ \ket{\bphi} := \ket{\phi_x}^{\otimes \secp^2 \cdot p^2}, \]
            which simply consists of copies of the prover's witness state.
            
            Compute $(\by, \brho) \gets \mathsf{C}.\Commit(\pk, \ket{\bphi})$ before sending $\by$. Let $\ell$ denote the number of qubits in $\ket{\bphi}$.
            \item[$\cV_3 \to \cP_3$:] Sample and send a random bit $b \in \{0, 1\}$.
            
            If $b = 0$:
            \begin{enumerate}
                \item $\cV_3 \to \cP_3$: Send a random bit $h \gets \{0,1\}$.
                \item $\cP_3 \to \cV_3$: Send $\bz \gets \Open(\brho, ([\ell], h^{\ell}))$.
                \item $\cV_3$: Compute $v = \Ver(\sk, \by, ([\ell], h^{\ell}), \bz)$ and accept if $v = 1$ and otherwise, reject.
            \end{enumerate}
            If $b = 1$:
            \begin{enumerate}
                \item $\cV_3 \to \cP_3$: Send random seeds $\bs_1, \bs_2 \gets \{0,1\}^{\secp}$.
                \item $\cP_3 \to \cV_3$: Compute $\bb = \PRG(\bs_1)|_{[1:\ell]} \in \{0,1\}^{\ell}$ and send the opening $\bz \gets \Open(\brho, ([\ell], \bb))$.
                \item $\cV_3$: Compute $\bb = \PRG(\bs_1)|_{[1:\ell]}$ and $r := \PRG(\bs_2) \in \{0, 1\}^{\poly(\ell)}$.
                
                Compute $u = \Ver(\sk, \by, ([\ell], \bb), \bz)$ and $\bv = \Out(\sk, \by, ([\ell], \bb), \bz)$. Accept if and only if $u = 1$ and $\V'_\MF(1^{\secp}, x, r, \bb, \bv) = 1$.
            \end{enumerate}
        \end{itemize}
    \end{protocol}
    \begin{proposition}
        Protocol \ref{protocol:classical_semi_succinct} is a constant-round classical $\QIA$ for $\calL$ with completeness $1-\negl(T(\secp))$ and soundness $1-\frac{1}{\secp^2}$ against $\poly(T(\secp))$-time adversaries. The prover runtime is $\poly(t_P)$, the verifier's verdict can be computed in time $\poly(t_P)$, and the verifier's messages are $\poly(\secp)$-sized.
    \end{proposition}
    \begin{proof}
        We first prove the complexity guarantees, followed by the completeness and soundness guarantees.        
        
        \paragraph{Complexity.} The fact that the prover's runtime is $\poly(t_P)$ and that the verifier's verctict function is computable in time $\poly(t_P)$ follows from the runtimes of the commitment scheme and the $\PRG$ computation, together with the assumption that $t_P \geq \secp$. The fact that the verifier's messages are $\poly(\secp)$-sized follows from the efficiency of the commitment scheme.
        
        \paragraph{Completeness.} By the $T$-security of $\PRG$, it suffices to swap $r$ with truly uniform randomness for $\V_{\MF}$ (as $\V_{\MF}$ runs in time $\poly(t_P) \leq \poly(T)$). Completeness then follows directly from the correctness of the commitment scheme and the completeness of $(P_2, V_2)$.
        
        \paragraph{Soundness.} Fix $P^{*}$ which runs in time $\poly(T)$, an input $x^{*} \notin \calL$, and an auxiliary state $\bsigma$, such that $P^{*}(x^{*}, \bsigma)$ is accepted with probability at least $1-\delta$, for $\delta < \frac{1}{\secp^2}$. We will use $P^{*}$ to construct a cheating prover $P^{**}$ for $(P_2, V_2)$ that runs in time $\poly(T)$ and succeeds with non-negligible probability in $T(\secp)$. The cheating prover $P^{**}$ behaves as follows:
        \begin{enumerate}
            \item Generate $(\pk, \sk) \gets \Gen(1^{\secp})$.
            \item Generate $(\by, \brho) \gets P^{*}(\pk, x^{*}, \bsigma)$.
            \item Use the extractor $\Ext$ from the commitment scheme to extract a state $\btau \gets \Ext^{P^{*}}(\sk, \by, \brho)$.
            \item Send $\btau$ to $V_2$.
        \end{enumerate}
        Note that $P^{*}$ runs in $\poly(T)$ time, and thus the extractor and $P^{**}$ run in $\poly(T)$ time.
    
        It must be the case that for each $h \in \{0, 1\}$, $P^{*}$ passes the test of the opening of the commitment to $h^{\ell}$ with probability at least $1-4\delta$; additionally $P^{*}$ must open to $\PRG(\bs_1)$ with probability at least $1-2\delta$ on average over random $\bs_1$. Thus, if we denote by 
            \[ \mathsf{Good} = \{\bs_1 \in \{0, 1\}^{\secp}: P^{*} \text{ is accepted w.p. $\geq 1-\secp\delta$ over $\bs_2$ when $V_3$ sends $\bs_1$}\}, \]
        then we have that $\Pr[\bs_1 \in \mathsf{Good}] \geq 1-\frac{2}{\secp}$ by a simple Markov argument.
    
        Therefore, by the binding property of the underlying commitment scheme, this means that for any basis $\bb = \PRG(\bs_1)|_{[1:\ell]}$ such that $\bs_1 \in \mathsf{Good}$, we have that
            \[ (\pk, \by, \bb, \bm_{\Real}) \stackrel{O(\sqrt{\secp\delta})}{\approx} (\pk, \by, \bb, \bm_{\Ideal}), \]
        where $\bm_{\Ideal}$ is the result of measuring $\btau$ in the basis $\bb$, and $\bm_{\Real}$ is the output corresponding to the opening of $P^{*}$ for $\by$ in the basis $\bb$. Thus, for any basis $\PRG(\bs_1)|_{[1:\ell]}$ where $\bs_1 \in \mathsf{Good}$, $\bm_{\Ideal}$ is accepted by $\V'_{\MF}$ with probability at least $1-\secp\delta-O(\sqrt{\secp\delta})$ over $\bs_2$.

        This means that $\V'_{\MF}$ accepts $\btau$ over pseudorandom $\bb$ and $r$ with probability at least
            \[ \left(1-\secp\delta-O\left(\sqrt{\secp\delta}\right)\right) \cdot \left(1-\frac{2}{\secp}\right) \geq 1-O\left(\frac{1}{\sqrt{\secp}}\right). \]

        But as $\V'_{\MF}$ runs in $\poly(t_P) = \poly(T)$ time, the $T$-security of $\PRG$ implies that $\V'_{\MF}$ actually accepts $\btau$ over \emph{random} $\bb$ and $r$ with probability at least
            \[ 1-O\left(\frac{1}{\sqrt{\secp}}\right)-\negl(T(\secp)) \gg \negl(T(\secp)), \]
       in contradiction to the soundness of $(P_2, V_2)$.
    \end{proof}

    \item Let $(P_4, V_4)$ be the $O(\secp^3 \cdot \log T) = O(\secp^4)$-fold sequential repetition of $(P_3, V_3)$.
    
    $(P_4, V_4)$ is a $T$-sound $O(\secp^4)$-round classical $\QIA$ with prover runtime $\poly(t_P)$ and verifier runtime $\poly(t_P)$ where the verifier's messages are $\poly(\secp)$-sized.
    
    \item Observing that the verifier's messages can also be computed obliviously to the prover's messages and in $\poly(\secp)$ time, we can apply the compiler from Theorem \ref{thm:BKL+22}, producing our final compiled protocol $(P', V')$.
\end{enumerate}
\end{proof}

\section{Acknowledgments}\label{sec:acknowledgments}
The authors would like to thank Anand Natarajan for many insightful discussions.

This material is based upon work supported by the Defense Advanced Research Projects Agency (DARPA) under Contract No. HR0011-25-C-0300. Any opinions, findings and conclusions or recommendations expressed in this material are those of the author(s) and do not necessarily reflect the views of the Defense Advanced Research Projects Agency (DARPA).

\bibliographystyle{alpha}
\bibliography{references.bib}

\appendix
\section{Proof of Theorem \ref{thm:XZ_tester}}\label{appendix:XZ_tester}
\begin{proof}[Proof of Theorem \ref{thm:XZ_tester}]
    We begin by introducing some notation from \cite{KSV02} that we use in our proof.
    
    \begin{definition}[\cite{KSV02}]
        For $T \geq 1$ and $t \in \{0, \ldots, T\}$, define $t_{\bone,T} \in \{0, 1\}^T$ to be the unary encoding
            \[ t_{\bone,T} = 1^t 0^{T-t}. \]
    \end{definition}
    
    \newcommand{\history}{\mathsf{history}}
    
    \begin{definition}[\cite{KSV02}]\label{def:history}
        Fix any quantum circuit $C$ that takes as input an $n$ qubit state and consists of $T$ gates $U_1, \ldots, U_T$. Then for any input state $\ket{\psi}$, we define the history state of $C$ on $\ket{\psi}$ to be 
            \[ \ket{\mathsf{history}_C(\psi)} = \frac{1}{\sqrt{T+1}} \sum_{t = 0}^T U_{\leq t} \ket{\psi} \otimes \ket{t_{\bone,T}},\]
        where $U_{\leq 0} = I$ and $U_{\leq t} = U_tU_{t-1} \ldots U_1$ for $1 \leq t \leq T$. Define the unitary $U_{\mathsf{history}_C}$ to be
            \[ U_{\mathsf{history}_C} = \sum_{t = 0}^T U_{\leq t} \otimes \ket{t_{\bone,T}}\bra{t_{\bone,T}}. \]
        Note that
            \[ U_{\mathsf{history}_C} \left(\ket{\psi} \otimes  \frac{1}{\sqrt{T+1}} \sum_{t = 0}^T \ket{t_{\bone,T}} \right) = \ket{\mathsf{history}_C(\psi)}. \]
    \end{definition}
    
    \begin{definition}
        A Hermitian operator $M$ is said to be \emph{$Y$-free} if the unique Pauli decomposition of $M$ has nonzero coefficients only for tensor products of $I$, $X$, and $Z$.
    \end{definition}

    As a first step, we prove the following lemma:
    \begin{lemma}[Derived from \cite{KSV02}] \label{lemma:history_qia} 
        Fix any function $t = t(|x|)$. Let $\calL$ be any language with a $\QMATIME[t]$ verifier $V$ that has completeness $c$, soundness $s$, and a complete set of real-valued witness states $\{\ket{\psi_x}\}_{x \in \calL}$.
        
        Then there exists a $\poly(t)$-time classical algorithm which on input $x$ outputs a Hamiltonian $H_x$ with that the following properties:
        \begin{enumerate}
            \item $H_x = \sum_{S} d_S S$ is a Hamiltonian which acts on $\poly(t)$ qubits, where each $d_S$ is a nonzero real number, and each $S$ is a tensor of the Pauli operators $X$, $Z$ and $I$ where at most 6 of them are not $I$. The number of coefficients $d_S$ is at most $\poly(t)$ and the sum of the absolute value of the coefficients satisfies $\sum_{S} |d_S| \leq  \poly(t)$. \label{item:Hpaulidecomp}
            \item For every $x \in \calL$, given $\ket{\psi_x}$ there is a $\poly(t)$-time computable state $\ket{\phi_x}$ such that
                \[ \bra{\phi_x} H_x \ket{\phi_x} \leq O\left(\frac{1-c}{\poly(t)}\right). \]
            \label{item:Hcomplete}
            \item For every $x \notin \calL$ and any state $\ket{\psi'_x}$,
                \[ \bra{\psi'_x} H_x \ket{\psi'_x} = \Omega\left(\frac{1-\sqrt{s}}{\poly(t)}\right). \]
            \label{item:Hsound}
        \end{enumerate}
    \end{lemma}
    The proof of Lemma \ref{lemma:history_qia} follows almost directly from Kitaev's circuit-to-Hamiltonian construction, from his proof of $\QMA$-completeness of the local Hamiltonian problem, as presented in~\cite{KSV02}. There, a similar theorem was proved; the main difference is that in their theorem, it was not required that the auxiliary state is real valued, and in Item \ref{item:Hpaulidecomp}, it was not required for the Hamiltonian to be $Y$-free (i.e., each $S$ is a local tensor of Pauli operators $X$, $Z$, $Y$, and $I$). 

    The proof of Lemma~\ref{lemma:history_qia} makes use of the following two claims.
    \begin{claim}\label{claim:y-free}
      The Hadamard gate $H$, Toffoli gate $CCZ$, and any diagonal Hermitian operator $\Delta$ can be written as a linear combination of tensor products of $I$, $X$, and $Z$.
    \end{claim}
    \begin{proof}[Proof of Claim \ref{claim:y-free}]
    For the Hadamard gate, write
        \[ H = \frac{1}{\sqrt{2}}X + \frac{1}{\sqrt{2}} Z. \]
        
    For the Toffoli gate, first observe that
        \[ \ket{11}\bra{11} = \frac{1}{4} (I-Z) \otimes (I - Z) = \frac{1}{4} (II - IZ - ZI + ZZ). \]
    Using the definition of the Toffoli gate
        \[ CCZ = (I - \ket{11}\bra{11}) \otimes I + \ket{11}\bra{11} \otimes Z, \]
    the conclusion follows.
          
    Any diagonal operator can be written as $\Delta = \sum_i d_i \ket{i}\bra{i}$, and each projector $\ket{i}\bra{i}$ has a Pauli decomposition consisting solely of tensor products of $I$ and $Z$ Paulis.
    \end{proof}
    \begin{claim}\label{claim:pauli-1norm}
      Let $M$ be a Hermitian operator that acts nontrivially on $k$ qubits, with operator norm $\|M \| > 0$. Then its Pauli decomposition $M =\sum_{P} c_P P$ satisfies the properties that the number of nonzero coefficients $c_P$ is at most $4^k$, and $\frac{1}{2^{k/2}} \|M\| \leq \sum_P |c_P| \leq 2^{k} \|M\|$.
    \end{claim}
    \begin{proof}[Proof of Claim \ref{claim:pauli-1norm}]
        Since $M$ acts nontrivially on $k$ qubits, we may write it as $M = N \otimes I$, where $N$ is a $k$-qubit operator with $\|N\| = \|M\|$. We will now argue about the Pauli decomposition of $N$, given by $N = \sum_{P} d_P P$. The conclusions about $M$ will follow because the Pauli decomposition of $M$ is simply $\sum_P d_P P \otimes I$. In particular, the bound on the number of nonzero Pauli coefficients follows immediately, since there are only $4^k$ Paulis acting on $k$ qubits.
        
        Recall that for an $n$-dimensional vector $v$, we have the inequalities $\sqrt{n} \|v\|_2 \geq  \|v\|_1 \geq \|v\|_2$. We will apply this to the vector $d$ of Pauli coefficients, which has dimension $4^k$. For this vector, 
        \begin{align*}
            \|d\|_1 &= \sum_P |d_P| \\
            \|d\|_2 &= \sqrt{\sum_{P} d_P^2 } = \sqrt{\sum_{P, P'}
                           d_P d_{P'} \frac{1}{2^k} \tr[PP']} \\
                    &= \frac{1}{2^{k/2}} \| M \|_2 .
        \end{align*}
        Now, further recall that for a $2^k \times 2^k$-dimensional matrix $M$, we have
            \[ 2^{k/2} \|M\| \geq \|M\|_2\geq  \|M\|. \]
        Putting these facts together, we have
        \begin{align*}
            \sum_P |d_P| &\geq \|d \|_2 = \frac{1}{2^{k/2}} \|M\|_2 \geq \frac{1}{2^{k/2}} \|M\|, \\
            \sum_P |d_P| &\leq \sqrt{4^k} \|d\|_2 = 2^{k/2} \|M\|_2 \leq 2^{k} \|M\|.
        \end{align*}
    \end{proof}
    
    \begin{proof}[Proof of Lemma \ref{lemma:history_qia}]
        Since our auxiliary states are real-valued, we can take the verifier circuit $V$ and convert it to a circuit that uses only Hadamard and Toffoli gates by Theorem \ref{thm:real-valued} with only polynomial overhead, negligible error, and the same auxiliary/witness state (with one additional ancilla qubit set to $\ket{0}$). Theorem \ref{thm:real-valued} implies that this new circuit $V'$ is in $\QPTIME[t]$ and has the same properties (up to negligible error).
    
        Denote the gate decomposition of $V'(x, \cdot)$ into Hadamard and Toffoli gates by $U_1, \ldots, U_T$. By assumption, $T = \poly(t)$ and $V'$ acts on $\ell := \poly(t)$ qubits, some of which are ancilla qubits. Let $\sfQ$ and $|\sfQ| \leq \poly(t)$ denote the register containing all such ancilla qubits and its size, respectively (recall that all ancillas are supposed to be initialized to $\ket{0}$).
        
        We start by defining a set of operators, which one can think of as acting on a unary-encoded ``clock'' register. These are defined in Equation~13.23 of~\cite{KSV02}, but we restate them here in a slightly different notation. Specifically, let us define the following matrices acting on $T$ qubits.
        \newcommand{\clk}{\mathsf{clock}}
        \begin{align*}
          \clk(\ket{0}\bra{0}) &= \ket{0}\bra{0} \otimes I^{\otimes (T - 1)} \\
          \clk(\ket{j}\bra{j}) &= I^{\otimes j-1} \otimes \ket{10}\bra{10} \otimes I^{ \otimes (T-j-1)}, \: \forall j \in \{1, \dots, T-1\} \\
          \clk(\ket{T}\bra{T}) &= I^{\otimes (T-1)} \otimes \ket{1}\bra{1} \\
          \clk(\ket{1}\bra{0}) &= \ket{10}\bra{00} \otimes I^{(T-2)} \\
          \clk(\ket{j}\bra{j-1}) &= I^{\otimes (j-2)} \otimes \ket{110}\bra{100} \otimes I^{\otimes(T - j - 1)},\: \forall j \in \{2, \dots, T-1\} \\
          \clk(\ket{T}\bra{T-1}) &= I^{\otimes(T-2)} \otimes \ket{11}\bra{10} \\
          \clk(\ket{j-1}\bra{j}) &= \clk(\ket{j}\bra{j-1})^\dagger, \: \forall j \in \{1, \dots, T \}.
        \end{align*}
        Observe that each of these matrices is at most $3$-local. Moreover, each one of these matrices is either diagonal, or of the form $\Delta \otimes \ket{1}\bra{0}$ or $\Delta \otimes \ket{0}\bra{1}$ up to a permutation of the qubits, where $\Delta$ is some diagonal matrix. 
        
        We now write the Hamiltonian. It will act on $\ell+T = \poly(t)$ qubits.
        \begin{align}
            H_{init} &= \sum_{i=1}^{|\sfQ|} I^{\otimes(\ell-|\sfQ|+i-1)} \otimes \ket{1}\bra{1} \otimes I^{\otimes(|\sfQ|-i)} \otimes \clk(\ket{0}\bra{0}) \\
            H_{clock} &= \sum_{j=1}^{T} I^{\otimes \ell} \otimes I^{\otimes(j-1)} \otimes \ket{01}\bra{01} \otimes I^{\otimes(T-j-1)} \\
            H_{prop,j} &= \frac{1}{2} \Big(-U_j \otimes \clk(\ket{j}\bra{j-1}) - U_j^\dagger \otimes \clk(\ket{j-1}\bra{j})  \nonumber \\
                    &\qquad + I \otimes (\clk(\ket{j}\bra{j}) + \clk(\ket{j-1}\bra{j-1})) \Big)  \label{eq:hpropj} \\
            H_{prop} &= \sum_{j=1}^{T} H_{prop,j} \\
            H_{final} &= I^{\ell-1} \otimes \ket{0}\bra{0} \otimes \clk(\ket{T}\bra{T}).
        \end{align}
        The Hamiltonian itself is
        \begin{equation}
          H_x = H_{prop} + H_{init} + H_{clock} + H_{final}.
        \end{equation}
        Note that by construction, $H_x \geq 0$, and $H_{init}, H_{prop, j}$, $H_{clock}$, and $H_{final}$ are all $6$-local. The total number of terms in $H_x$ is $(|\sfQ|+1)+T+T+1 = \poly(t)$. Since $H_{init}$, $H_{clock}$, and $H_{final}$ are orthogonal projectors, they have operator norm exactly equal to 1. To see that $H_{prop}$ has operator norm 2, consider the conjugation of each term $H_{prop, j}$ by the unitary $W = \sum_{t = 0}^T U_{\leq t} \otimes \ket{t_{\bone, T}}\bra{t_{\bone, T}}$:
            \[ W^{\dagger} H_{prop, j} W = I \otimes \frac{1}{2}(\ket{j-1_{\bone, T}}\bra{j-1_{\bone, T}}+\ket{j_{\bone, T}}\bra{j_{\bone, T}}-\ket{j-1_{\bone, T}}\bra{j_{\bone, T}}-\ket{j_{\bone, T}}\bra{j-1_{\bone, T}}) := I \otimes E_j. \]
        Thus, $W^{\dagger} H_{prop} W = I \otimes \sum_j E_j := I \otimes E$, and this matrix $E$ has largest eigenvalue (and hence operator norm) bounded by 2 (see \cite{KSV02}).
    
        To prove Item~\ref{item:Hpaulidecomp}, we first show that each term of $H_x$ is $Y$-free.
        \begin{itemize}
            \item Each term of $H_{init}$ is diagonal, and hence $H_{init}$ is $Y$-free by Lemma \ref{claim:y-free}.
            \item Similarly, $H_{final}$ and each term of $H_{clock}$ is diagonal, and hence they are $Y$-free.
            \item For each $j$, the terms of $H_{prop,j}$ can be split into two cases. For the second two terms in \cref{eq:hpropj}, these terms are diagonal, so they are $Y$-free. For the first two, we must use the conditions about the gates in $C$ in the hypothesis of the theorem. Specifically, we know that every gate $U_j$ in $C$ is either a Hadamard or Toffoli gate (and hence Hermitian). Thus, the first two terms in \cref{eq:hpropj} are proportional to
                \[ U_j \otimes (\clk(\ket{j}\bra{j-1}) + \clk(\ket{j-1}\bra{j})). \]
            Now, we use the observation made earlier that $\clk(\ket{j}\bra{j-1})$ can be written after permutation of the qubits as $\Delta \otimes \ket{1}\bra{0}$. This means that, after permutation of the qubits, the first two terms in \cref{eq:hpropj} are proportional to
                \[ U_j \otimes \Delta \otimes (\ket{1}\bra{0} + \ket{0}\bra{1}) = U_j \otimes \Delta \otimes X. \]
            This is the tensor product of three $Y$-free Hermitian matrices and so is also a $Y$-free Hermitian matrix.
        \end{itemize}
    
        Thus, each term $h_i$ in $H_x$ can be written as a sum $h_i = \sum_{P} c_{i,P} P$ where the Paulis $P$ are all tensor products of $I, X, Z$ only. Additionally, by \cref{claim:pauli-1norm}, for each $i$, the number of nonzero coefficients $c_{i,P}$ is at most $4^6$, and $1/2^3 \leq \sum_P|c_{i,P}| \leq 2^7$. To write the whole Hamiltonian, we simply take the sum of the Pauli decompositions of each term, without combining ``duplicate'' Paulis: that is, if the same Pauli $P$ appears in the decomposition for $h_i$ and $h_j$, we included it twice with coefficients $c_{i,P}$ and $c_{j,P}$ in our full decomposition. This gives us a decomposition
            \[ H_x = \sum_{S} d_S S, \]
        where each $S$ is a tensor product of $I, X, Z$, and $\sum_{S} |d_S| = \poly(t)$. This proves \cref{item:Hpaulidecomp}. Each term of $H_x$ is computable by simply determining the appropriate clock matrices and (possibly) the corresponding Hadamard or Toffoli gate (which admit efficiently computable Pauli decompositions). Since $V' \in \QPTIME[t]$, it is not hard to see that the final decomposition of $H_x$ is also computable in time $\poly(t)$.
        
        In addition, by taking $\ket{\phi_x} := \history_{V'(x, \cdot)}(\ket{0}_{\sfQ} \otimes \ket{\psi_x})$, we observe that $\ket{\phi_x}$ can be computed efficiently given $\ket{\psi_x}$ by the efficiency of the history state construction. \Cref{item:Hcomplete} follows by observing that such a history state is a ground state of $H_{init}$ and $H_{prop}$, and is penalized by $H_{final}$ only when, conditioned on having measured the final clock register, the output of the circuit is not 1, which occurs with probability $(1-c)/(T+1) = O((1-c)/\poly(t))$. 
    
        \Cref{item:Hsound} follows exactly as in \cite{KSV02, AN02} by the soundness of the $\QMATIME[t]$ verifier $V'(x, \cdot)$ (as $H_{init}$ penalizes inputs which do not have $\ket{0}$ in the ancilla registers).
    \end{proof}
    
    We can now instantiate the Morimae-Fitzsimons protocol (see Theorem \ref{thm:MF16}) with the Hamiltonian $H_x$ associated with Lemma \ref{lemma:history_qia}. We assume without loss of generality that $\S_{\MF}$ always produces the same number of nonempty indices (equal to the locality of the underlying Hamiltonian) by padding appropriately.
    
    Since $H_x$ acts on $\poly(t)$ qubits and has $\poly(t)$ local terms, $\S_{\MF}$ runs in $\poly(t)$ time and thus uses at most $\poly(t)$ bits of randomness. Thus, it is simple to modify $\S_{\MF}(H_x)$ to also take as input randomness $r \gets \{0, 1\}^{\poly(t)}$ which it will use to deterministically return a basis to measure in; we will refer to this ``derandomized'' version by $\S'_{\MF}(H_x; r)$.
    
    All that is left to do is to modify the Morimae-Fitzsimons protocol (as in \cite{ACGH20}) to make it instance-independent. This is done by additionally sampling a uniformly random basis $\bb \gets \{0, 1\}^{\poly(t)}$ to measure the state in and verifying only consistent measurements with $\S'_{\MF}(H_x; r)$. That is, given measurement outcomes $\bm$ produced by measuring $\brho$ in the basis $\bb$ and randomness $r$, the verifier $\V_{\MF}$ behaves as follows:
    \begin{itemize}
        \item Compute $\vec{b} \gets \S'_{\MF}(H_x; r)$. If $\vec{b}$ and $\bb$ are not consistent (i.e. $\exists i: \vec{b}_i \neq \bot \land \vec{b}_i \neq \bb_i$), then accept immediately. Else, output $\D_{\MF}(H_x, \vec{b}, \bm)$.
    \end{itemize}
    
    Since $H_x$ is 6-local, for any $r$, $\S'_{\MF}(H_x; r)$ returns a fixed basis $\vec{b}$ with exactly 6 non-empty indices. Thus, a uniformly random $\bb$ will be consistent with $\vec{b}$ with probability $\frac{1}{64}$. Thus, we have that
    \begin{align*}
        \Pr_{\bb \gets \{0, 1\}^{\poly(t)}, r \gets \{0, 1\}^{\poly(t)}}[\V_\MF(x, r, \bb, \Measure(\bb, \ket{\psi})) = 1] &= \frac{63}{64}+\frac{1}{64}\left(\frac{1}{2}-\frac{\bra{\psi} H_x \ket{\psi}}{\sum_S 2|d_S|}\right) \\
        &= \frac{127}{128}-\frac{\bra{\psi} H_x \ket{\psi}}{128\sum_S|d_S|}.
    \end{align*}
    
    When $x \in \calL$, taking $\ket{\phi_x}$ to be the $\poly(t)$-computable state guaranteed by Lemma \ref{lemma:history_qia}, we have that
        \[ \Pr_{\bb, r}[\V_\MF(x, r, \bb, \Measure(\bb, \ket{\phi_x})) = 1] \geq \frac{127}{128}-O\left(\frac{1-c}{\poly(t)}\right), \]
    by \cref{item:Hpaulidecomp} and \cref{item:Hcomplete} of Lemma \ref{lemma:history_qia}.
    
    When $x \notin \calL$, for all states $\brho_x$, we have that
        \[ \Pr_{\bb, r}[\V_\MF(x, r, \bb, \Measure(\bb, \brho_x)) = 1] \leq \frac{127}{128}-\Omega\left(\frac{1-\sqrt{s}}{\poly(t)}\right), \]
    by \cref{item:Hpaulidecomp} and \cref{item:Hsound} of Lemma \ref{lemma:history_qia}.
\end{proof}
\end{document}